\documentclass{acm_proc_article-sp}

\usepackage{graphicx}
\usepackage{epsfig}
\usepackage{amssymb}
\usepackage{amsmath}
\usepackage{url}
\usepackage{enumitem}
\usepackage[numbers]{natbib}
\usepackage{notoccite}

\begin{document}

\title{The Effects of Data Quality on the Analysis \\ of Corporate Board Interlock Networks} 

\numberofauthors{2} 

\author{ \alignauthor
Javier Garcia-Bernardo\\
 \affaddr{CORPNET, University of Amsterdam}\\
 \email{garcia@uva.nl}
\alignauthor
Frank W. Takes\\
 \affaddr{CORPNET, University of Amsterdam}\\
 \affaddr{LIACS, Leiden University}\\
 \email{takes@uva.nl}
}

\date{15 November 2016}

% TODO: we should stress somewhere that having an idea of the completeness is really important in board interlock studies, and is often neglected. JGB: Do we have a reference?

\maketitle
\begin{abstract}

Nowadays, social networks of ever increasing size are studied by researchers from a range of disciplines. 
The data underlying these networks is often automatically gathered from API's, websites or existing databases. 
As a result, the quality of this data is typically not manually validated, and the resulting networks may be based on false, biased or incomplete data. 
In this paper, we investigate the effect of data quality issues on the analysis of large networks.
We focus on the global board interlock network, in which nodes represent firms across the globe, and edges model social ties between firms -- shared board members holding a position at both firms. 
First, we demonstrate how we can automatically assess the completeness of a large dataset of 160 million firms, in which data is missing not at random. 
%This allows us to assess the completeness of the data for each country in the global network.
Second, we present a novel method to increase the accuracy of the entries in our data.
By comparing the expected and empirical characteristics of the resulting network topology, we develop a technique that automatically prunes and merges duplicate nodes and edges.
Third, we use a case study of the board interlock network of Sweden to show how poor quality data results in incorrect network topologies, biased centrality values and abnormal influence spread under a well-known diffusion model. 
Finally, we demonstrate how our data quality assessment  methods help restore the correct network structure, ultimately allowing us to derive meaningful and correct results from analyzing the network. 

\end{abstract}

% FWT: nevermind this for the initial submission
%\category{H.4}{Computer Applications}{Social and behavioral sciences}[]


%\keywords{[keywords]} % NOT required for Proceedings

%%%%%%%%%%%%%%%%%%%%%%%%%%%%%%%%%%%%%%%%%%%%%%%%%%%%%%%%%

\section{Introduction}
\label{sec:introduction}

% introduce data quality: completeness and accuracy
Over the past few decades, the amount of digital information has been doubling roughly every two years. 
At the same time, there is an ongoing and prevailing desire to extract meaningful knowledge from this data. 
Although many knowledge discovery methods and techniques are scalable to larger volumes of data, ``big data''~\cite{ward2013undefined} has the significant and largely unaddressed problem of ``veracity''. 
This refers to the fact that the explosion in the amount of available data has resulted in a situation in which researchers can no longer manually validate the \emph{quality} of their data~\cite{wand1996anchoring}. 
Data quality most dominantly relates to questions of \emph{completeness} (what part of the data do we have, and what part do we miss?) and \emph{accuracy} (is the data that we have correct and suitable for answering our particular domain questions?). 
Here we set out to assess how these issues can be addressed in the context of (social) network analysis. 

% corporate networks
In this paper we focus on so-called \emph{corporate networks}, in which ties represent particular relationships between corporations.
Ties in corporate networks can be based on a variety of relationships between firms, including 
trade~\cite{wilhite2001bilateral}, 
borrowing and lending of money~\cite{battiston2016complexity}, ownership~\cite{vitali2011network}, 
or as we will analyze in this paper: shared board members. 
In these networks of \emph{interlocking directorates}, also referred to as \emph{board interlock networks}, a node represents a firm and an edge between two firms denotes that these firms share at least one board member or director. 
An example of a board interlock network is given in Figure~\ref{fig:examplenetwork}.
Board interlocks are common practice in today's corporate world, and over the past century, social scientists have extensively studied the causes and consequences of board interlocks. 
See for example the excellent overview given in~\cite{mizruchi1996interlocks}, where Mizruchi discusses how interlocks relate to collusion, monitoring (e.g., banks keeping an eye on firms they invested in), legitimacy (attracting board members with a particular reputation in a certain area that is of importance to the firm), individual career advancement and social cohesion (social ties among the upper class). 
Previous research has established that networks of interlocking directorates facilitate the spread of governance routines and practices, the exchange of resources, communication and the dissemination of new ideas~\cite{burris2005interlocking,davis1991agents}. 
%This information diffusion typically happens at the monthly board of directors' meetings,
%where all directors in a firm sit together.
Since a significant number of directors % fix suggested by eelke 16-11-2016
 has positions at two or more firms, the board meetings of these firms connect the majority of big businesses in the world.
For instance, Davis~\cite{davis1991agents} discusses how the majority of the corporate elite would rapidly be infected by a contagious disease as a result of the small world property of the network of interlocking directorates. 

% something about analyzing board interlock networks
Corporate networks have interesting topological characteristics common to real-world networks, including a fat tailed degree distribution, the emergence of a giant component and very low pairwise distances between nodes~\cite{battiston2004statistical}. 
Researchers have thus applied established social network analysis methods and techniques~\cite{scott2012social} to corporate networks. 
Community detection has been used to understand the geographical dimension of the structure of these networks~\cite{heemskerk2016corporate,vitali2013community} and centrality measures provide insight into powerful firms and countries in the network~\cite{takesheemskerk2016}. 
%Indeed, methods for analyzing social networks have been successfully applied to corporate board interlock networks to understand the influence of network effects on firms and their behavior as societal and economic actors. 
% interlocking directorates network properties, problem with big data

Initially, social scientists studied only small networks of interlocking directorates, typically based on a few hundred firms and their relationships. 
Researchers carefully double-checked the data they manually gathered from annual reports of the companies involved. 
Nowadays, large databases on corporations are provided by commercial corporate information providers such as Orbis, BoardEx, ThomsonOne and Bloomberg, including information on their financial performance and board composition. 
The availability of these databases with millions of firms allows our board interlock networks to be automatically constructed from the available firm and board member data. 
Here, we focus on a dataset extracted from Orbis, 
a large information provider that gathers data from country registers across the world, and then makes this data available through one database (for details, see Section~\ref{sec:data}). 
The sheer volume of this contemporary big corporate network data (BCND, see~\cite{heemskerk2016big}) means that it is no longer possible to manually check each firm, let alone their board composition, for correctness. 
However, the quality of our data is diverse across countries and country registers.
Indeed, the problem of data quality (completeness and accuracy) comes into play here.

% completeness details and example
The first data quality issue, \emph{completeness}, is not %always 
necessarily problematic, for example when we have a dataset in which data is missing completely at random (MCAR), or when missing values are directly correlated with a known variable (MAR).
We find that in our dataset information about some attributes (e.g., number of employees) is correlated with other attributes with better availability -- it is MAR. 
However, often it is not the attributes but the companies themselves that are not present in the data, meaning data is missing \emph{not} at random (MNAR). 
This may result in severe problems, because if non-random parts of the data are missing, we can no longer consider it a reliable sample and derive meaningful results for the system represented by the dataset as a whole. 
For example, if we blindly use the data provided by Orbis to compare countries, we observe that the average Mexican firm is larger than the average firm in the United States. 
In turns out that this is due to lower data quality in Mexico, where many small companies are not included in the data, thus increasing the average size of a company. 
If we want to derive meaningful and actionable insights from the board interlock network of these corporations, we need to know exactly which firms we are missing. 

% accuracy details and example
A second problem, \emph{accuracy}, is that we have no prior indication of whether there are duplicates in our data. 
These duplicates can be accidental, for example as a result of entity resolution errors introduced when the data was gathered by the information providers, but they can also be the result of administrative reasons. 
Firms often organize themselves in multiple legal entities and parent-subsidiary constructions to facilitate the autonomy of local branches, departments or, most often, for certain legal or financial benefits. 
We want to disregard these types of administrative firms and their relationships from the network, such that we are able to derive meaningful and representative results from a network in which a node represents one distinct and autonomous firm, and an edge represents a true board interlock between these firms, facilitating the type of social relation between firms discussed in our short survey of board interlock research provided above. 

% constribution and remainder of paper
The two data quality issues of completeness and accuracy of corporate datasets obviously affect the topology of board interlock networks. 
Missing, spurious or duplicate nodes or edges affect %local neighborhoods of nodes (firms) in the network, but also 
the analysis of the network as well, for example in terms of topological properties, centrality and the diffusion of information~\cite{kossinets2006effects}. 
In this paper, we address these data quality problems in corporate network data.
For the issue of completeness we present in Section~\ref{sec:completeness} a new quality assessment method to analyze which part of the global firm data can be used to ensure meaningful results. 
For the aspect of accuracy, Section~\ref{sec:accuracy} proposes a data correction technique that uses network characteristics to filter spurious firms and connections in the corporate network. 
To demonstrate how results change from fuzzy to more meaningful as data quality issues are addressed, we analyze in Section~\ref{sec:results} their effect on the network topology, centrality and influence spread in the network of interlocking directorates of Sweden. 
Finally, we show in Section~\ref{sec:conclusion} that the proposed method of assessing and correcting the quality of corporate network data is essential to ensure that results from analyzing the network are not a mere product of systematic biases in the data, but instead provide actually meaningful insights. 
Finally in Section~\ref{sec:futurework} we discuss extensions of the proposed methods to other commonly studied (social) networks. 
%I we want to we could say "Finally, we show in that our results are generalizable to other domains, and that the proposed method of assessing and correcting the quality [...]"

\begin{figure}[!bt]
	\centering
	\includegraphics[width=1.05\linewidth]{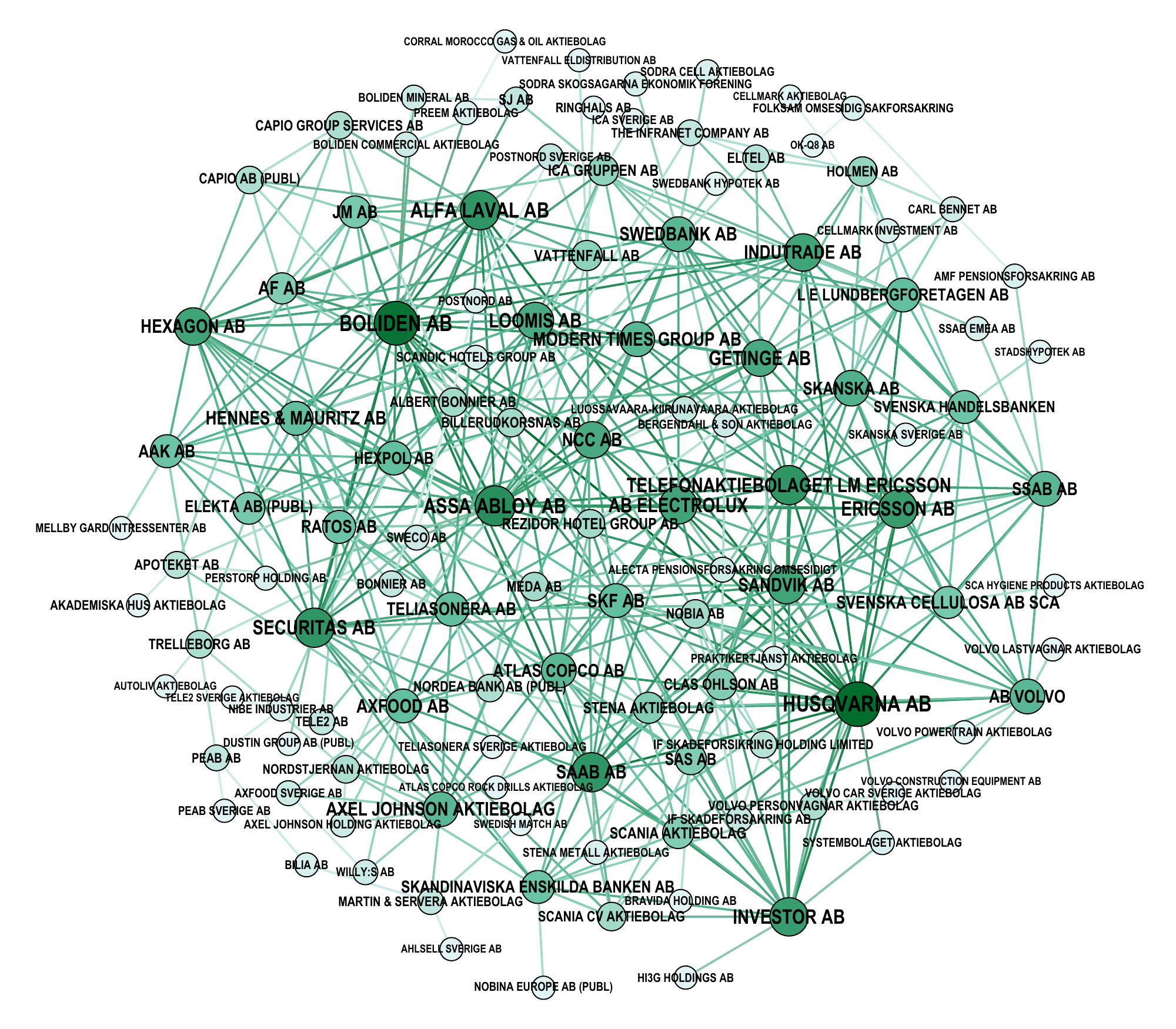}
	\caption{Sample of the Swedish network of interlocking directorates. Based on largest 120 firms in terms of revenue, connected through 422 interlocks. Color and size are proportional to node degree.} 
\end{figure}\label{fig:examplenetwork}

%%%%%%%%%%%%%%%%%%%%%%%%%%%%%%%%%%%%%%%%%%%%%%%%%%%%%%%%%

\section{Data} \label{sec:data}

This section outlines how we collected our dataset, followed by a description of how we constructed the board interlock network that is analyzed in the experimental sections. 

We started by creating a snapshot of the Orbis database\footnote{Orbis, Bureau van Dijk, \url{http://orbis.bvdinfo.com}} in November 2015,
including all 160 million active companies (no branches, foreign entities or business marked as `single location') and all 90 million directors holding a position at these companies.
For each company we extracted its country, operating revenue, number of employees and global ultimate owner (GUO). % when available. 
The GUO of a company is a controlling shareholder (directly or indirectly owning at least $50\%$ in shares) that is not owned by any other company.
For each director we extracted all top executive (chief and director) affiliations in the set of firms, obtaining 65 million current positions. 
This dataset is the object of study in our analysis of completeness in Section~\ref{sec:completeness}. 

Next, we constructed the Swedish firm-to-firm board interlock network, 
where only Swedish firms are considered, and companies are connected if they share one or more directors.
The resulting Swedish network is composed of 260,611 companies and 1,269,560 edges.
Although the network also contains many uninteresting small components of at most a handful of nodes, we focus on its giant component in which the lion share of economic activity in Sweden takes place. 
This connected undirected graph contains a total of 94,496 nodes and 1,050,907 edges.
In Section~\ref{sec:accuracy} and Section~\ref{sec:results} we assess the quality in terms of accuracy of this network dataset, investigating its topology and the results of applying methods such as centrality and influence spread models. 

\begin{figure*}[htb]
	\centering
	\includegraphics[width=\textwidth]{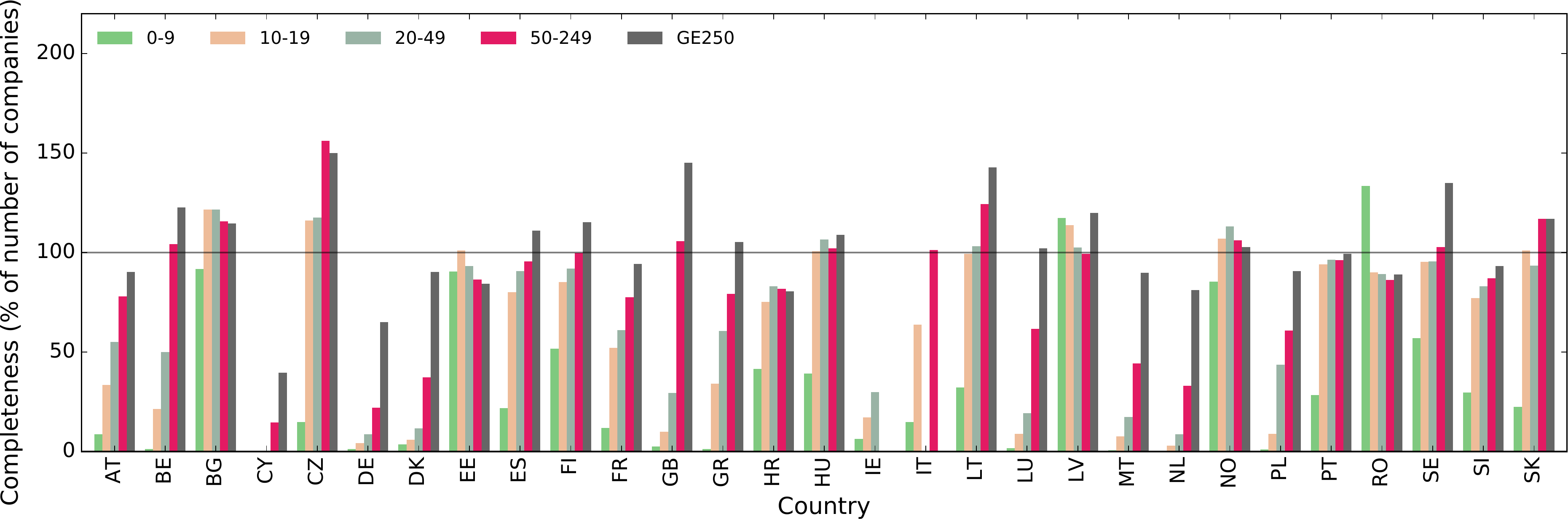}
	\caption{Percentage of companies present in our dataset as a function of the number of employees working at these companies: <10, 10-19, 20-49, 50-249 and >250 employees.} 
 \label{fig:eurostat}
\end{figure*}

\begin{table*}[htb]
\centering
\begin{tabular}{l|l|l|l}
Variable code  & Variable name   & Mean effect & \% Models \\ \hline
NY.GDP.PCAP.KD\_2013 & GDP per capita (constant 2005 US\$) & 0.263048 & 23.2 \\
SH.XPD.PUBL.GX.ZS\_2013 & Health expenditure, public (\% of government ex... & 0.220553 & 25.7 \\
logAG.YLD.CREL.KG\_2013 & Cereal yield (kg per hectare)  & 0.207313 & 24.8 \\
logIC.TAX.DURS\_2013 & Time to prepare and pay taxes (hours) & -0.200198 & 24.0 \\
AG.YLD.CREL.KG\_2013 & Cereal yield (kg per hectare)  & 0.198041 & 22.4 \\
logTX.VAL.MRCH.R5.ZS\_2013 & Merchandise exports to developing economies in... & 0.191862 & 22.8 \\
logIC.EXP.DURS\_2013 & Time to export (days)  & -0.184246 & 21.4 \\
BM.TRF.PWKR.CD.DT\_2013 & Personal remittances, paid (current US\$) & 0.179447 & 27.5 \\
SP.RUR.TOTL.ZG\_2013 & Rural population growth (annual \%) & 0.174098 & 27.8 \\
IC.TAX.TOTL.CP.ZS\_2013 & Total tax rate (\% of commercial profits) & -0.169697 & 33.4
\end{tabular}
\caption{Predictors for the mean company revenue in a country ($\hat{R}$), and average coefficient of the support vector (mean effect), 
obtained using the Python \textit{sklearn} (\texttt{http://scikit-learn.org}) package with a linear kernel and the default penalty parameter $C=1$.
The rightmost columns shows the percentage of simulations in which the variable was among the top 10 predictors. 
Only the indicators present in more than 20\% of the repetitions were considered.
If the variable code is preceded with 'log', then this variable was log-transformed. %\vspace{-8mm}
}
 \label{table:indicators}
\end{table*} 

\section{Methodology} \label{sec:methods}

The two major components of data quality, namely data completeness and accuracy, are discussed and addressed in the two subsections below, specifically in the context of corporate data and the corporate board interlock network.

\subsection{Completeness} \label{sec:completeness} 

Ensuring proper data analysis results requires unbiased representative samples of a population.
Unbiased representative samples are obtained when the distributions of each variable in the sample match the distributions in the real population.
However, the distributions in the population are generally unknown, i.e., our dataset has data missing not at random (MNAR). 
In the case of our corporate data, we usually either lack information about the distribution of company sizes in a given country, 
or have only aggregated data -- e.g., the number of small, medium and large companies available.

Here, we propose to use aggregated data on segments of the data to test the completeness of the full dataset. 
For instance, Figure~\ref{fig:eurostat} shows the coverage of our dataset by category (in terms of number of employees) as reported by the European statistics bureau\footnote{EUROSTAT, \url{http://ec.europa.eu/eurostat}}.
Some countries (e.g., Norway (NO), Sweden (SE), Finland (FI) and Estonia (EE)) have very complete information (all bars are close to the horizontal line). 
However, some countries have relatively bad data quality (e.g., Poland (PL), the Netherlands (NL) and Germany (DE)).
Moreover, we see that countries with bad data quality usually have good information for large companies and bad information for small companies.
%Although European countries generally report the number of companies and its combined revenue in rough categories, many other countries do not.
%If we want to ensure good results, we need to analyze the effect of these systematic biases in our results.
In order to understand the magnitude of the bias we first find the precise relationship between quality and macroeconomic measures, using countries with available aggregated data.
We then use this relation to extrapolate to countries without information. 
In this section, we apply our method to a dataset on corporations, where aggregated information at country level is available. 
However, our method can be applied for all datasets where aggregated information on one or more of the individual' attributes is available, given that we can estimate the underlying distribution.
In Section \ref{sec:futurework} we discuss in detail the generalizability of our approach.

We start by modeling the company size (in terms of its operating revenue) as a lognormal distribution.
Lognormal distributions arise naturally in multiplicative processes, where the (in this case revenue) growth 
between time $t-1$ and $t$, denoted 
$R_t - R_{t-1}$, depends on the combinations of a series of factors $F_k$ (for example, random fluctuations, type of business, population density, income of customers, etc.): $R_t -R_{t-1} = F_k\cdot R_{t-1}$. 
This model is known as the Gibrat's law~\cite{gibrat1931inegalites}.
The idea here is that the size (represented by revenue) of companies varies in size according to $F_k$. 
After a certain total time $t$ (corresponding to the present time) it holds that 
\begin{equation}
R_t = R_0\cdot\prod_{k=1}^t{(1+F_k)},
\end{equation} 
and
\begin{equation}
\log{R_t} = \log{R_0} + \sum_{k=1}^t{\log{(1+F_k)}}.
\end{equation}
When using short time scales the effect of the factors is small, and $\log{(1+F_k)} \approx F_k$:
\begin{equation}
\log{R_t} = \log{R_0} + \sum_{k=1}^t{F_k}.
\end{equation}
Assuming that $F_k$ are independent and identically distributed (i.i.d.) 
we obtain that $R_t$ is lognormally distributed (central limit theorem)~\cite{Mitzenmacher}. 
Moreover, a small change in the model, namely that the number of time iterations varies for different companies,
produces power-law tails~\cite{Mitzenmacher,montroll1983maximum}, which we observe in our data, as we show at the end of this section.
It is worth noting that the combinations of the factors within $F_k$ are likely to follow lognormal distributions themselves. 
For instance, personal income follows lognormal distributions with power-law tails~\cite{Mitzenmacher,montroll1983maximum} and population densities have long tails~\cite{Mitzenmacher}.

\begin{figure}[ht]
	\centering
	\includegraphics[width=0.5\textwidth]{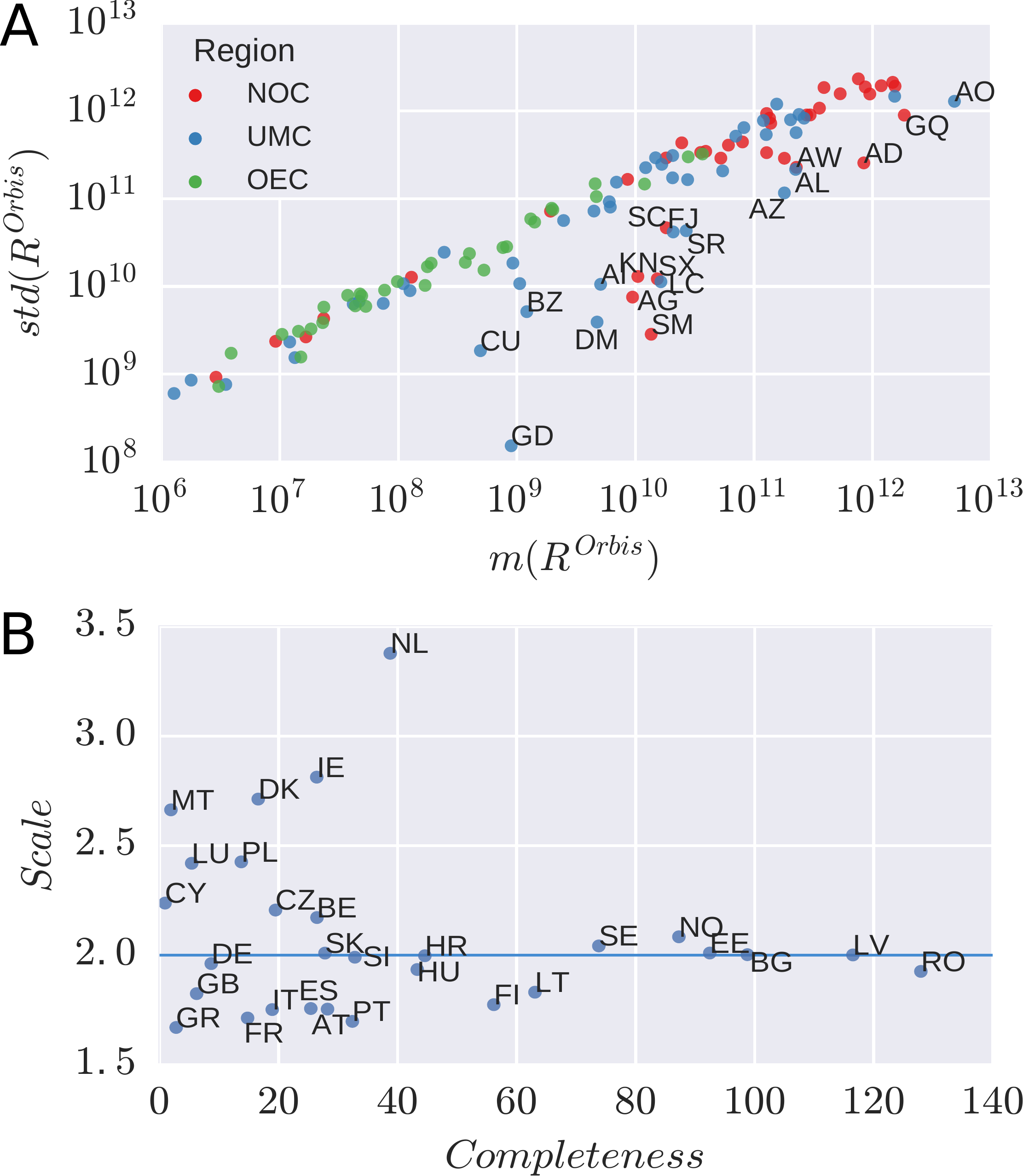}
 \caption{(A) Standard deviation vs. mean company revenue by country in high income OECD (OEC), high income nonOECD (NOC) and upper middle income (UMC) countries. 
 (B) Maximum likelihood estimator of $\sigma$ vs. percentage of companies (with respect to Eurostat) present in our dataset.}
 \label{fig:mean}
 \end{figure}

Either if $F_k$ are i.i.d. or if they exhibit long tails, revenue distributions can be well fitted by lognormal distributions.
%http://leo.ugr.es/pgm2012/proceedings/eproceedings/cobb_approximating.pdf
%http://www.jstor.org.proxy.uba.uva.nl:2048/stable/2296055?sid=primo&origin=crossref&seq=4#page_scan_tab_contents
Lognormal distributions are characterized by two parameters: scale $\sigma$ and location $\mu$.
Importantly, the standard deviation $s$ is proportional to the mean $m$: 
\begin{equation}
\log{(\mathit{s})} = \log{(\mathit{m})} + \frac{1}{2}\log{\left(e^{\sigma^2}-1\right)}
\end{equation}
When the means and standard deviations of the revenue of all companies in a country are plotted against each other, 
the values for every country with enough data lie in a straight line (see Figure~\ref{fig:mean}A), 
indicating that all countries share the same $\sigma$.
%The relation is even stronger when the mean and standard deviation of the market capitalization in a country are plotted together, since data on market capitalization has good quality. % Javier, is (any proof of) this sentence necessary to mention?
Since $\sigma$ clusters around $2.0$ in countries with known better quality (Figure~\ref{fig:mean}B), 
we fixed $\sigma = 2.0$. 

\begin{figure}[ht]
	\centering
	\includegraphics[width=0.5\textwidth]{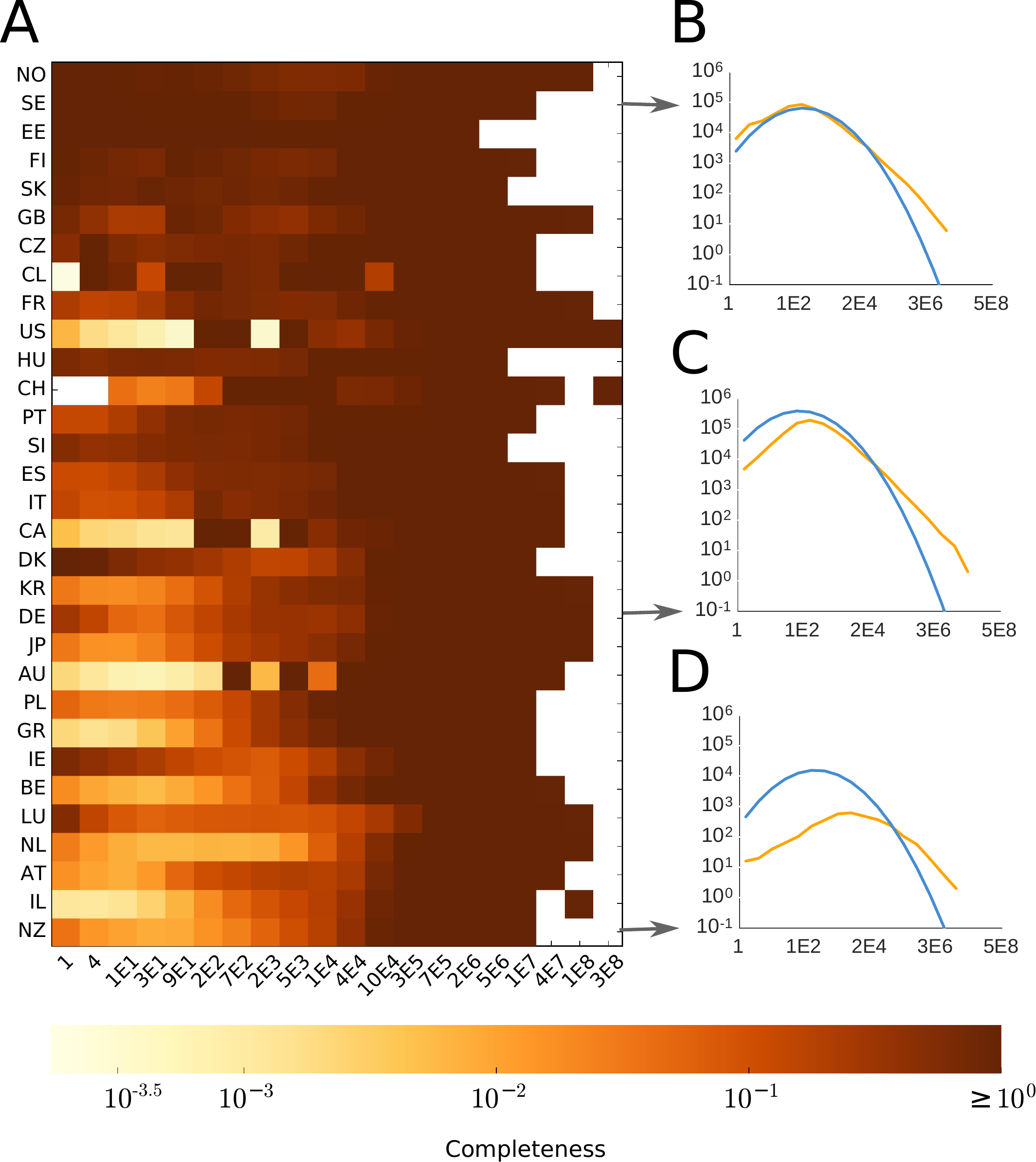}
 \caption{(A) Fraction of companies present by revenue category. (B--D) Distribution of company revenue in our database (orange) and distribution estimated using $\mu$ and $C$ (blue), for (B) Sweden, (C) Germany and (D) New Zealand.}
 \label{fig:imshow}
 \end{figure}

Although this approach allows us to fix the scale and find the maximum likelihood $\mu$ parameter of the distributions, $\mu$ would be biased by data quality. 
In particular, we know that rich countries have better quality and a better reporting of small companies, which lowers the $\mu$ parameters. 
However, rich countries have larger companies, with increases the $\mu$ parameter. This confounding effect can be disentangled by using country-level aggregated data. 
In order to find the real $\mu$ parameter we obtained the number of companies and combined revenue for all OECD countries\footnote{OECD SSIS\_BSC\_ISIC4, Cat. 05\_82\_LESS\_K}. All currencies were converted to USD using exchanges rates on March 25th '16, adding India and completing missing information in Canada, Australia and the United States from their statistics bureaus\footnote{The aggregated revenue of companies in India and Canada was estimated from the GDP of the country using the linear relationship between $\log{GDP}$ and $\log{(\mathit{total\ revenue})}$ of the other countries ($R^2$ = 0.975)}. 

We then estimated the logarithm of the theoretical mean of the revenue distribution $\log{(\hat{R})}$ by using world development indicators (WDI)\footnote{WorldBank WDI, \url{http://wdi.worldbank.org}}. 
Since $\log{(\hat{R})} = \mu + \sigma^2/2$ and $\sigma=2.0$, estimating $\log{(\hat{R})}$ is equivalent to estimating $\mu$.
In order to prevent overfitting, we fitted 1000 linear models using support vector regression on random samples %of the countries 
containing 75\% of the countries. We iteratively dropped the worst indicator until we found a core of 10 indicators. 
As shown in Table~\ref{table:indicators}, the main indicator is GDP per capita. 
In fact, this indicator alone can explain 72\% of the variability in OECD countries (82\% if Norway is excluded).
In general, larger income taxes and bureaucracy times are correlated with smaller companies, while larger productivity, GDP per capita and personal remittances paid are correlated with larger companies.

After estimating the revenue average we confirmed that it closely matches the average extracted from the OECD data ($R^2 = 0.91$). 
We then used this relationship between $\log{(\hat{R})}$ and WDI indicators to estimate $\log{(\hat{R})}$ for all countries in the world.
Next, we hypothesized that companies are added to the source database in decreasing order of revenue.
If this would be true, there would be a quasi-linear country-level relationship between the estimated average using WDI indicators $(\hat{R})$, and the product of the percentage of companies present in the dataset ($C$) and the average revenue in our dataset $(R^{\mathit{Obs}})$.
We found that this was indeed the case ($R^2 = 0.82$), and the three variables are related as follows:
\begin{equation}
\log{\mathit{C}} = -1.3855 - 0.954 \log{(R^{\mathit{Obs}})} + 1.1120 \log{(\hat{R})}.
\end{equation} 
Thus, for any given country we can now calculate the theoretical average (and $\mu$) using WDI indicators,
and the completeness using the theoretical average and our dataset average.
Finally, we estimated $\mu$ and therewith the completeness $C$ for all countries.
We compared the distribution of revenues in our dataset with the expected distribution using $\mu$ and $C$ (Figure~\ref{fig:imshow}A).
As expected, countries with good quality in Figure~\ref{fig:eurostat} also have good quality in Figure~\ref{fig:imshow}A.
Canada, Australia and the US have sharp peaks at specific revenue bins, which are caused by lax reporting requirement in those countries.
Moreover, similarly to income distributions~\cite{Mitzenmacher,montroll1983maximum}, 
we observe that revenue distributions are well-described by lognormal distributions with power-law tails for the $\sim 1\%$ larger companies (Figure~\ref{fig:imshow}B--D). We conclude from Figure~\ref{fig:imshow}A that data quality in terms of completeness for large firms is generally good, while data quality for small firms depends on the considered country.

\begin{figure*}[ht] 
	\centering
	\includegraphics[width=\textwidth]{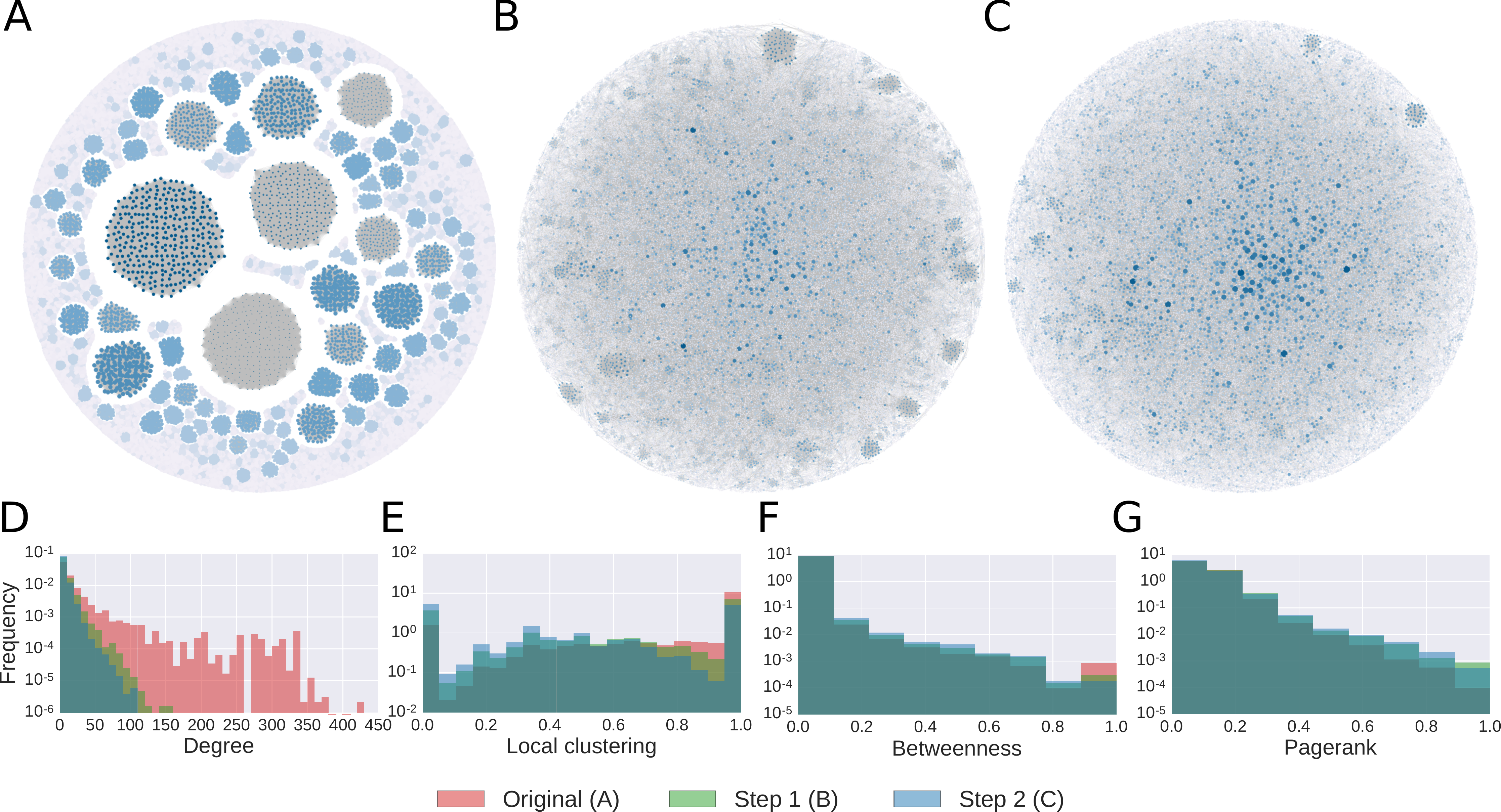}
 \caption{%Accuracy fixes and centrality distributions. 
 (A--C) Swedish network visualizations using the ForceAtlas2 algorithm~\cite{jacomy2014forceatlas2} %,bastian2009gephi} 
 with stronger gravity to highlight clusters with high local clustering. Color indicates PageRank centrality. (A) Original network, (B) Network after Step 1, (C) Network after Step 2. (D--G) Comparison of distributions of network measures for the three networks: (D) Degree, (E) Local clustering, (F) Betweenness and (G) PageRank.}
 \label{fig:distributions}
\end{figure*}

%%%%%%%%%%%%%%%%%%%%%%%%%%%%%%%%%%%%%%%%%%%%%%%%%%%%

\subsection{Accuracy} \label{sec:accuracy}

In addition to the problem of missing data, we may have duplicated data, which relates to the aspect of \emph{accuracy}. 
Our data was gathered from different sources and then combined, a process which is error prone as a result of a lack of unique identifiers for firms and directors. 
In case of corporate data, companies can be reported several times or split into several parts for administrative or financial reasons, as discussed in Section~\ref{sec:introduction}. 
%In order to create dataset in which a node represents one unique company 
To counter this, we propose two fixing steps:

\begin{enumerate}[wide, labelwidth=!, labelindent=2em]%[leftmargin=*]
\item An \emph{a posteriori network construction step} in which we merge nodes (companies) with exactly the same board of directors and the same GUO (or having no GUO).
\item An \emph{a posteriori topology-based correction step} in which we merge all nodes (companies) sharing a similar board and similar \emph{position} in the network. 
To compute the similarity of boards of two companies, we used the ratio of shared directors to total unique directors (Jaccard similarity~\cite{crandall2008feedback}). 
We then clustered all companies to
% TODO Javier: what is "complete linkage"?JGB: It's an agglomeration method where the distance between clusters is calculated using the farthest nodes from each cluster (that way you guarantee that each node within a cluster is within a distance).
obtain sets of firms where each pair of companies shared at least half of the boards of directors (using complete linkage clustering and a threshold for the Jaccard similarity of 0.5 to guarantee this condition).
% Javier: explain why 0.5? (or why it is OK and does not have to be reported on) JGB: I could run some experiments to see how much it change if necessary, it's a restrictive case where companies 1 and 2 may be merged if their boars are 1: directors A & B, 2: director A (1 out of 2 unique directors), but not if there are 1: A & B 2: A & C (1 out of 3).
Then, we further clustered the groups based on their local position in terms of a) degree, b) average neighbor degree, c) local clustering coefficient and d) average neighbor local clustering.
All nodes with a similar board and with these four properties within 80\% of each other are then merged together.
While other threshold values are possible, our results are robust to variations in this parameter: changing the parameter to 90\% or 70\% changed the final number of nodes only by 2.2\% and 2.4\%, respectively.

\end{enumerate}
%In Section~\ref{sec:results} we demonstrate the benefits of these steps in analyzing the corporate board interlock network.
\section{Results} \label{sec:results}
%Data quality can impair the results from social network research.

After analyzing the completeness of our data (Section~\ref{sec:completeness}), 
we have chosen to use the network of Sweden to demonstrate the effect of the two accuracy fixes proposed in Section~\ref{sec:accuracy}. 
As shown in Figure~\ref{fig:imshow}, Sweden (SE) has high quality across all company sizes,
which will prevent confounding interactions between completeness and accuracy.
We investigate the effect on network topology in Section~\ref{sec:results_topology}, centrality in Section~\ref{sec:results_centrality} and finally diffusion in Section~\ref{sec:results_sir}.
%in two examples.
%In the first example, we use several centrality measures to rank companies by importance, and compare our ranking with the ranking by number of employees and market capitalization.
%In the second example, we use a simple SIR model to show how diffusion models can produce biased resulted if we use data of bad quality.
%To demonstrate the effect of 
 %(Figure~\ref{fig:distributions}). % not sure why this fig is referenced here.
 
 \begin{table}[b]
\setlength{\tabcolsep}{0.5em}

 \begin{tabular}{l|rrrrrrr}
~ & Nodes & Edges & Density & $\overline{deg}$ & ${CC}$ & $\overline{d}$ \\ 
\hline
 Original & 94496 & 1050907 & 0.0119\% & 22.2  & 0.93 & 7.78  \\
 Step 1 & 60904 & 225887 & 0.0061\% & 7.4   & 0.57 & 7.94  \\
 Step 2 & 50733 & 139173 & 0.0054\% & 5.5  & 0.41 & 8.02  \\
 \end{tabular}
	\caption{Topological properties of the Swedish board interlock network.} 
 \label{table:SE_measures}
\end{table}

\subsection{Topology} 
\label{sec:results_topology}

The original Swedish board interlock network has 94,496 nodes and 1,050,907 edges in the largest component (Section~\ref{sec:data}), 
which is reduced to 50,733 nodes and 139,173 edges after the two steps (Section~\ref{sec:accuracy}). 
Table~\ref{table:SE_measures} shows the number of nodes (firms), edges (board interlocks), % fix suggested by eelke 16-11-2016
density, average degree $\overline{deg}$, 
 graph clustering coefficient $CC$ and average node-to-node distance $\overline{d}$ (see~\cite{scott2012social} for definitions of these common network metrics). 

As expected given that we are merging clusters with high local clustering, the average degree and local clustering are reduced. 
Indeed, an average degree around $5$ or $6$ is far more realistic in board interlock networks than $22$. 
The power of our approach is further reflected in the network visualizations.
Figure~\ref{fig:distributions}A--C shows the original Swedish network, and the network after Steps 1 and 2 described in Section~\ref{sec:accuracy}.
While most of the edges in the original network were intra-corporate administrative ties,
the successive steps were able to filter out such clusters and provide a representative final network topology.

\begin{figure*}[!tb]
	\centering
	\includegraphics[width=\textwidth]{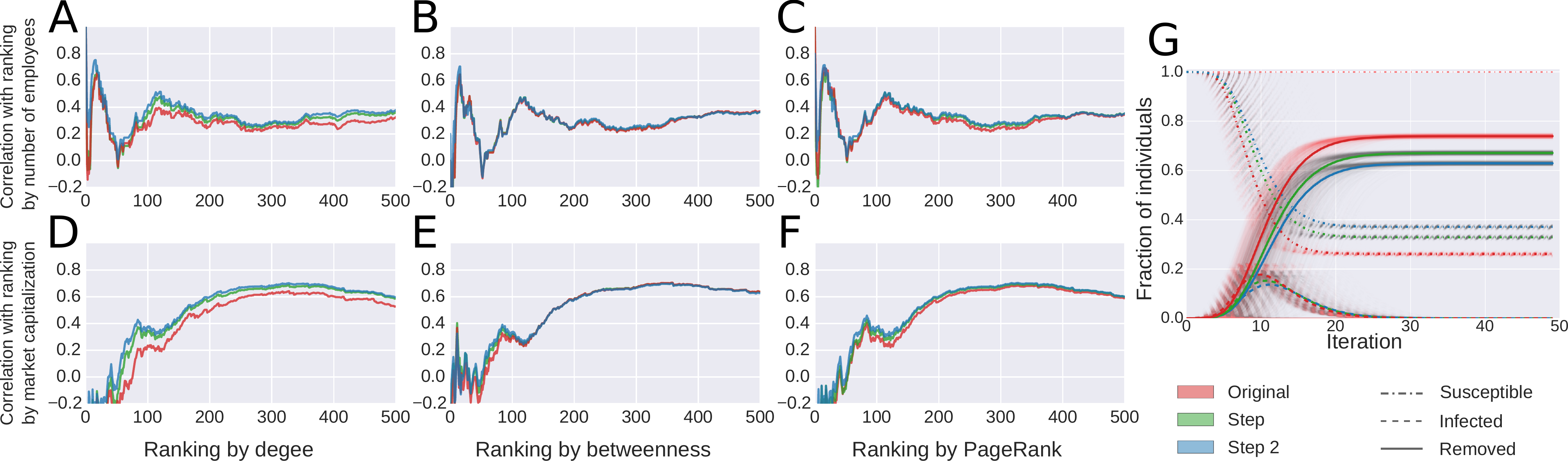}
 \caption{%[PNG placeholder] 
 (A--F) Correlation between ranking given by number of employees (A--C) or market capitalization (D--F), and degree, betweenness and PageRank centrality.
 (G) Fraction of individuals in the Susceptible, Infected and Recovered states. %All plots show results for the original data, and the data after step 1 and 2.
 }
 \label{fig:sir}
 \end{figure*}

\subsection{Centrality}
 \label{sec:results_centrality}
 
A typical step in social network analysis is to find the most prominent actors in the network, which is done using centrality measures.
In this example, we choose three common measures:
\textit{degree centrality}, in which the importance of each node is proportional to its number of connections;
\textit{PageRank centrality}, in which the importance depends on the sum of the PageRank of your neighbors damped by a factor;
and \textit{betweenness centrality}, that relates the importance of a node to the relative number of shortest paths going through the node. 
For an overview of work on the use of centrality measures in board interlock networks, see~\cite{takesheemskerk2016}. 

Figure~\ref{fig:distributions}D--G shows how local clustering and three centrality measures -- degree, betweenness and PageRank -- are affected by our data quality fixes.
In general, our filtering mechanism smooths the distribution of centrality measures.
While the degree distribution of the original data exhibits a fat tail (Figure~\ref{fig:distributions}D), with some companies linked to up to 500 other companies, this was not the case for the data after Step 1 and 2. 
This is largely due to the reduction of clusters with high local clustering (Figure~\ref{fig:distributions}E)
Importantly, the distributions of betweenness and PageRank (Figure~\ref{fig:distributions}F--G) are only minimally disturbed, showing a reduction of the number of nodes with very high betweenness, likely due to a reduction of clusters of high degree nodes.

Next, we tested if our data quality fixes improved our ability to obtain correct and thus actionable insights from a centrality analysis. 
For this, we analyzed the correlation between the ranking of Swedish companies by number of employees and market capitalization with the ranking of companies given by degree, by PageRank and by betweenness centrality (Figure~\ref{fig:sir}).
We found that each correction step increased the correlation between the ranking based on economic measures and the ranking based on network measures.
This was especially the case for degree centrality (Figure~\ref{fig:sir}AD), 
but only to a very small extent for betweenness centrality (Figure~\ref{fig:sir}BE).
Moreover, we found that the rankings obtained by all three network measures are comparable, 
which indicates that central nodes have more connections (degree), 
link to other important firms (PageRank),
and act as bridges between distant firms (betweenness).
Finally, we found that the ranking given by market capitalization closely matches the ranking given by centrality (Spearman's rank correlation of 0.6 for the top 200-500 companies),
showing that larger companies have also more central roles in the network.

\subsection{Diffusion model}
 \label{sec:results_sir}
 
As board interlock networks are assumed to play a big role in the diffusion of information, we tested how accuracy can affect the results of a SIR (susceptible, infected, recovered) model.
These models are commonly applied to model information diffusion,
as for example in the case of information among investors or technology diffusion between companies~\cite{geroski2000models,shiller1989survey}. 
To illustrate the effects of data quality, we have chosen to apply a simple SIR model:
starting with one randomly infected node, at each iteration of the algorithm, each infected node infects their neighbors with probability 0.5, and recovers (becoming immune) with probability 0.3.
Figure~\ref{fig:sir}G shows the results of running 1000 simulations. 

As we can observe, the process dies off in 19\%, 27\% and 37\% of the simulation runs for respectively the original data and the data after Steps 1 and 2.
Second, the final equilibrium differs.
Removing the clusters with high degree nodes slows down the spreading cascade and the percentage of nodes infected during the process,
from 74\% of infected nodes using the original network to 67\% using the post Step 1 network and 63\% using the post Step 2 network.
These results are likely due to the larger number of high degree nodes in the original network, which allows for more possibilities of infection,
and to the `reservoir' effect of clusters -- the infection quickly spreads within the clusters, infecting neighbors for a longer time until all nodes in the cluster recover.

%%%%%%%%%%%%%%%%%%%%%%%%%%%%%%%%%%%%%%%%%%%%%

\section{Conclusions} \label{sec:conclusion}

The experiments in this paper demonstrate how insights and predictions obtained from social network analysis can be biased if the original data is of poor quality. 
If we want to ensure that network analysis results are correct, we must determine the quality of the underlying data and where needed correct for it.
In this paper we used corporate board interlock networks to investigate the effect of data quality. 

The underlying company data is collected in a distributed fashion in different countries.
Since the missing information is not distributed at random,
%, consisting of financial indicators and information about boards of directors, is collected in a distributed fashion, with one or more data agencies per country, and is finally merged in the Orbis database from which we extracted our data.
%Moreover, different countries have different registering requirements and ability to gather company information.
%Thus, the missing information is not distributed at random,
assessing the \emph{completeness} of this data is challenging. 
To address this problem, we used aggregated data from the OECD
%.Stat 
%on the sum of firm revenues and total number of firms in a country. 
%We then find an empirical relationship between a country's economic macro-indicators and the average revenue of the companies in that country. 
and fitted lognormal distributions to the revenue of the companies of each country. 
%Next, we used the average revenue of the companies to fit a lognormal distribution, 
This enables us to calculate the number of missing companies for different revenue ranges, allowing for a thorough assessment of how much data we are missing in each country. 

The second problem addressed in this paper dealt with the \emph{accuracy} of the data. 
This originates from the fact that data comes from multiple merged data sources, and from the inherent way in which companies organize themselves. 
In order to not overestimate the importance of for example small companies, we proposed a method to merge corporate structures and remove spurious nodes.
%For instance, in big corporate network data we usually encounter several instances of a company coming from different sources, with similar information.
For this, we suggest two a posteriori solutions to ``correct'' the network topology, where we merge together companies with similar boards of directors and similar
structural positions in the network. 

In our experiments, we investigated the effect of data quality by visualizing the networks to show how the original biased network becomes a ``clean'' network.
First, we demonstrated how our correction approach fixed the overall network topology. 
Second, we show that the correlation between the most central nodes in the network and the nodes with higher market capitalization increases for the networks that were processed by our approach. 
Third, we show how a simple SIR model demonstrates that bad data quality network data produces significant distortions in the results. 

\section{Generalizability \& future \\ work}
\label{sec:futurework}

Although we have shown the applicability of our method in board interlock networks, the principles could be used across different networks.  The completeness assessment can be applied to each network in which we know the distribution of one of the variables in the dataset, and aggregated data on segments of the data is available.
%For instance, we can understand the completeness of a sample of academic papers by looking at their number of citations, since they follow a lognormal distribution with constant $\sigma$ and journal-specific $\mu$ ~\cite{radicchi2008universality}.
This makes our approach applicable to the analysis of (co-)citation networks, for example to investigate if the considered data sample is biased toward highly impact publications.
We could use the average number of citations to fit lognormal distributions for different journals~\cite{radicchi2008universality}, and use those distributions to assess the completeness of the sample.
Similarly, inter-event times in social communication (for instance emailing, tweeting or messaging) can also be well fitted by lognormal distributions~\cite{doerr2013lognormal},
whose parameters may depend on the characteristics of the person.
If we have a sample of the messages, we can assess its bias using the distribution of inter-event times.

Furthermore, the accuracy fixes that we introduced here can be applied when networks are compiled from multiple sources and have not been merged correctly.
For instance, consider combining data on people from two different sources (e.g., Linkedin, AngelList and CVs) based on their name. 
Since the affiliations in Linkedin, AngelList and the CV are correlated, it is likely possible to use our method to identify unique people. 

While our paper focuses
on completeness and accuracy of nodes, future work can use
similar principles to assess the completeness and accuracy 
of edges in the network. 
Finally, extensions to dynamic networks in which timestamp attributes are present, could be investigated.

\subsection*{Acknowledgments}
This research is part of the CORPNET project (\url{http://corpnet.uva.nl}) at the University of Amsterdam. 
The project has received funding from the European Research Council (ERC) under the European Union's Horizon 2020 research and innovation programme (grant agreement 638946).

\bibliographystyle{spmpsci}      % mathematics and physical sciences
\bibliography{sigproc} % sigproc.bib is the name of the Bibliography in this case
% and remember to run:
% ACM needs 'a single self-contained file'!
%

\balancecolumns

% That's all folks!

\end{document}